\documentclass[11pt,a4paper]{article}
\usepackage{jheppub}
\usepackage{comment}
\renewcommand\d{\partial}

\newcommand\Tr{\mathop{\mathrm{Tr}}}
\newcommand{\tab}{\hspace*{2em}}

\begin{document}

\preprint{EFI-13-36}
\title{Effective field theory for fluids: Hall viscosity from a 
Wess-Zumino-Witten term}
\author{Michael Geracie and Dam Thanh Son}
\affiliation{Kadanoff Center for Theoretical Physics 
and Department of Physics, University of Chicago, S Ellis Ave, Chicago, IL 60637, USA}
\emailAdd{mgeracie@uchicago.edu, dtson@uchicago.edu}

\abstract{We propose an effective action that describes a relativistic
  fluid with Hall viscosity.  The construction involves a
  Wess-Zumino-Witten term that exists only in (2+1) spacetime
  dimensions.  We note that this formalism can accommodate only a Hall
  viscosity which is a homogeneous function of the entropy and
  particle number densities of degree one.}


\maketitle

\section{Introduction}

Hydrodynamics provides a universal description of the long wavelength
dynamics of finite-temperature interacting systems.  It can be
considered an effective theory organized in a derivative expansion.  To
leading order one has a dissipationless theory (ideal hydrodynamics),
and effects of dissipation enter at the next-to-leading order
(first-order hydrodynamics).  The dissipative effects are parameterized
by kinetic coefficients (viscosities, thermal conductivity etc.)

Recently, it has been found that new, dissipationless terms are
possible in first-order hydrodynamics.  In 3+1 dimensions, these
effects are related to the anomalies of the underlying quantum field
theory.  In 2+1 dimensions, one such possible term involves the Hall
viscosity~\cite{Avron:1995fg}.  In a gapped quantum Hall fluid the
Hall viscosity has a topological nature, it is related to the shift
and the angular momentum density.  In a gapless fluid such a connection
is not expected to hold.  Holographic models typically lead to a Hall
viscosity which is not related to the angular momentum density,
although there are indications of a relationship in holographic
$p_x+ip_y$ superfluids near the phase transition~\cite{Son:2013xra}.
 
Since hydrodynamics, including Hall viscosity, is a dissipationless
theory, it is natural to ask whether a description based on an action
is possible.  This question has been addressed before, but no term has
been identified with the Hall viscosity.  The goal of this paper is to
show that such a term exists.  However, the Hall viscosity in this
description cannot be an arbitrary function of entropy density $s$ and
the particle number density $n$, but must have the form of $s f(n/s)$.
In particular, for an uncharged plasma, the Lagrangian description
corresponds to a constant ratio of the Hall viscosity and the entropy
density.

\section{Review of effective field theory formulation of fluid dynamics}

We begin with a brief review of an effective field theory (EFT)
approach to fluid dynamics (for details see
Refs.~\cite{Dubovsky:2005xd,Dubovsky:2011sj}).  We will at first
concentrate on the simplest case of a fluid without conserved charges,
where the fundamental degree of freedom in the EFT description is a
set of scalars $\phi^I(x,t)$, $I = 1, \dots d$ mapping spatial slices
to a $d$ dimensional internal space $\mathcal{F}$. We will set $d=2$
for the entirety of this paper.  One can interpret $\phi^I$ as
Lagrangian coordinates, labeling which fluid particle occupies the
point $x$ at the time $t$ (such an interpretation, however, should
not be taken literally.)  We seek to write down an action functional
of $\phi^I$ and its derivatives, invariant under shifts $\phi^I \to
\phi^I+a^I$ and area preserving diffeomorphisms,
\begin{equation}
	\phi^I \rightarrow \tilde \phi^I (\phi ), \quad
	\det (\partial_I \tilde \phi^J) = 1. \label{sym}
\end{equation}
Invariance with respect to
area preserving diffeomorphisms in $\phi^I$ space
is what tells us we're working with a fluid, not a solid.

We define, following Ref.~\cite{Dubovsky:2005xd}, a scalar
\begin{equation}
  s = \sqrt{\det \partial_\mu \phi^I \partial^\mu \phi^J } 
  = \frac{1}{\sqrt{-g}} \det \partial_i \phi^I ( 1 - \dot x^2 )^{1/2}
\end{equation}
and a unit timelike vector
\begin{equation}
	u^\mu = \frac{1}{2 s} \varepsilon^{\mu \alpha \beta} 
   \epsilon_{I J} \partial_\alpha \phi^I \partial_\beta \phi^J,
\end{equation}
where $\varepsilon^{\mu\nu\lambda}$ is the completely antisymmetric
tensor  $\varepsilon^{012} = 1/\sqrt{-g}$.
The vector $u^\mu$ is tangent to curves of constant $\phi^I$: 
\begin{align}
	u^\mu\partial_\mu \phi^I = 0. \label{fluid velocity}
\end{align}
The current
\begin{equation}
   s^\mu = s u^\mu =\frac{1}{2} \varepsilon^{\mu \alpha \beta} 
   \epsilon_{I J} \partial_\alpha \phi^I \partial_\beta \phi^J
	\label{entropy current}
\end{equation}
is then identically conserved $\nabla_\mu s^\mu = 0$, independent of
equations of motion and is identified with the entropy
current and $s$ with the entropy density~\cite{Dubovsky:2005xd}.

Due to the shift symmetry the fields $\phi^I$ always appear in the
Lagrangian with at least one derivative so in the construction of the
effective action we use a power-counting scheme in which
$\partial_\mu\phi^I$ is counted as having no spatial derivatives,
$\d_\mu\d_\nu\phi^I$ as having one derivative, etc.  To leading order
the only possible term in the action is
\begin{equation}\label{S-ideal}
   S = - \int\! d^3 x\, \sqrt{-g}\, \epsilon (s),
\end{equation}
where $s$ is an arbitrary function of one variable.  Under metric
variations, $\delta s = \frac12 s P_{\mu \nu} \delta g^{\mu \nu}$, and
the stress-energy tensor computed from~(\ref{S-ideal}) is
\begin{align}
  T_{\mu \nu} = \epsilon u_\mu u_\nu +  P_{\mu \nu} P,
\end{align}
where $P= s\d_s \epsilon - \epsilon$ and $P_{\mu \nu} = g_{\mu \nu} +
u_\mu u_\nu$ is the spatial projector.  This is exactly the
stress-energy tensor of a perfect fluid if one identifies
$\epsilon(s)$ as the energy density as a function of the entropy
density and its Legendre transform $P$ as the pressure.

\section{Wess-Zumino-Witten term}

Now we turn our attention to first-order corrections to hydrodynamics.
Previous attempts~\cite{Nicolis:2011ey,Haehl:2013kra} to include
parity-breaking terms have not produced nontrivial transport
coefficients: the stress contribution of all local, first order terms
that may be added to the Lagrangian may be removed by some suitable
field redefinition.  We show that previous analyses miss one term
which corresponds precisely to the Hall viscosity.  This term has the
form of a Wess-Zumino-Witten term, and our construction is very similar
to that of Ref.~\cite{Maciejko:2013dia}.

To begin, we define an induced metric on the space of comoving fluid
particles $\phi^I$,
\begin{align}
	G^{IJ} = s^{-1} \partial_\mu \phi^I \partial^\mu \phi^J.
\end{align}
The inclusion of the factor $s^{-1}$ implies that $\det G=1$.  Note
that $G^{IJ}$ transforms as a tensor under area-preserving
diffeomorphism.  We denote the inverse of $G^{IJ}$ by $G_{IJ}$ and the
volume form by $\epsilon_{IJ}$.  Since the metric is unimodular,
$\epsilon_{12} = 1$.  In what follows, we will often use matrix
notation in which we denote $G_{IJ}$ as $G$, $G^{IJ}$ as $G^{-1}$, and
both $\epsilon_{IJ}$ and $\epsilon^{IJ}$ by $\epsilon$ (context should
make clear whether we are working with raised or lowered indices).

We follow the general philosophy behind the construction of
Wess-Zumino-Witten (WZW) terms~\cite{Witten}.  We first concentrate on
one particular fluid element, parameterized by some values of the
comoving coordinates $\phi^I$.  Points on the world line of this fluid
element can parameterized further by a parameter $\tau$.  For example,
$\tau$ can be (but does not have to be) chosen as the proper time.  At
the chosen values of $\phi^I$, the field $G_{IJ}(\tau)$ defines a map
from the world line to the space of unimodular matrices, which we call
$\mathbb H^2$.

There is a two-form on $\mathbb H^2$ which is
invariant under area preserving diffeomorphisms,
\begin{align}\label{omega-def}
  \omega = \frac{1}{2} \text{Tr} 
     \big( \epsilon\, d G\, G^{-1} \wedge d G \big).
\end{align}
This form is clearly closed, as it is a two-form on a two-dimensional
space. Since $\mathbb H^2$ is contractible, the form is also exact
so there exists a one-form $\alpha$ so that $\omega=d\alpha$.  We
define the WZW action for each fluid volume element to be the integral of
$\alpha$.  Note that $\alpha$ is defined only up to a differential of
a zero-form, but a replacement $\alpha\to\alpha+d\beta$ merely changes the
Lagrangian by a total derivative.  For the whole fluid, we sum up
contributions from each of the fluid elements,
\begin{equation}
  S_{\rm WZW} = f  \int\! d^2\phi\! \int\!\alpha, \label{WZW}
\end{equation}
where $f$ is some constant. The second integral is taken along the worldline
of each fluid element.

The form $\omega$ can be interpreted as the volume form of a natural
geometry on $\mathbb H^2$.  To make our discussion more concrete, we
follow Ref.~\cite{Maciejko:2013dia} and parameterize $\mathbb H^2$ via
\begin{align}
	G_{IJ} = 
	\begin{pmatrix}
		T+X & Y \\
		Y & T-X
	\end{pmatrix}.
	\label{plane}
\end{align}
The condition $\det G=1$ implies $T^2-X^2-Y^2=1$, and the positive
definiteness of $G$ implies $T > 0$.  Thus the space of all $G_{IJ}$
is the two-dimensional hyperbolic plane which caries a natural metric induced by
the Minkowski metric of the $(T,X,Y)$ space.  To see
that $\omega$ is the volume form, we can choose to use the coordinates 
$( Q, \varphi )$
\begin{equation}
	X = \sinh Q \cos \varphi, \tab
	Y = \sinh Q \sin \varphi, \tab
	T = \cosh Q,
\end{equation}
with $0 \leq Q \leq \infty$ and $0 \leq \varphi < 2 \pi$, and find by
explicit calculation from Eq.~(\ref{omega-def}) that $- \frac{1}{2} \omega=\sinh Q\, 
dQ\wedge d\varphi$, which is the volume form on the hyperbolic plane.

One convenient way to write the WZW action is by
extending the metric $G_{IJ}$ to an extra dimension $u$,
$G_{IJ}=G_{IJ}( \phi ,\tau ,u)$, so that it interpolates between a
reference metric, say $\delta_{IJ}$, at $u=0$ to the physical metric
at $u=1$,
\begin{equation}
	G_{IJ} ( \phi , \tau , 0 ) = \delta_{IJ}, \tab
	G_{IJ} ( \phi , \tau, 1 ) = G_{IJ} ( \phi , \tau ).
\end{equation}
The WZW term then can be written as
\begin{equation}
	S_{\rm WZW} =  f\! \int\! d^2 \phi\, d\tau\, du\, \text{Tr} 
\big( \epsilon \partial_u G G^{-1} \partial_\tau G \big).
\end{equation}
Instead of writing the action as an integral over the fluid
(Lagrangian) coordinates $(\tau,\phi^I)$, one can also write it in the
physical (Eulerian) coordinates $x^\mu$
\begin{equation}
   S_{\rm WZW} = f\! \int\! d^3x\, du\,\sqrt {-g}\, s\, \text{Tr} 
\big( \epsilon \partial_u G G^{-1} u^\mu\d_\mu G \big), \label{WZW coord}
\end{equation}
where $\partial_\tau = u^\mu \partial_\mu$ and $G= G(\phi, \tau, u)$.
This makes explicit that $S_{WZW}$ is first order in derivatives, and
should be included in our expansion. Though not manifest in this form,
it is insensitive to the details of the interpolation and the choice
of reference metric since in the end it is just a rewriting of
Eq.~(\ref{WZW}).

Let's check this explicitly and vary the interpolation $\delta
G$. Since the interpolation must begin at a fixed point, and end at
the physical metric, we have $\delta G |_{u=1} = 0$ and $\partial_\tau
\delta G |_{u=0} = 0$. Under such a change
\begin{equation}
   \delta S_{\rm WZW} = f\!\int\! d^2\phi\, d\tau\, du  \, \text{Tr} 
   ( \epsilon \partial_u \delta G G^{-1} \partial_\tau G 
   + \epsilon \partial_u G \delta 
G^{-1} \partial_\tau G + \epsilon \partial_u G G^{-1} \partial_\tau \delta G).
\end{equation}
The second term vanishes and using integration by parts on $\tau$, the
first and third can be combined as a total $u$-derivative
\begin{align}
   \delta S_{\rm WZW} = f \!\int\! d^2 \phi\, d\tau\, du \, \partial_u \big(
   \text{Tr} ( \epsilon \delta G G^{-1} \partial_\tau G ) \big) = 0 .
  \label{total_derivative}
\end{align}

\section{Stress-energy tensor}

We now vary this with respect to the metric to obtain its stress
contribution $\Delta T_{\mu \nu} = - 2\delta S_{\rm WZW}/\delta g
  ^{\mu \nu}$. The variation is the total $u$-derivative (\ref{total_derivative}) which
integrates to
\begin{equation}
  \delta S = f\! \int\! d^2 \phi\, d \tau \, \Tr ( \epsilon \delta G^{-1} 
G \partial_\tau G^{-1} ),
\end{equation}
where $G$ is the physical ($u=1$) metric. Using the following identities
\begin{align}
   \epsilon_{IJ} &= - s u^\mu \varepsilon_{\mu \nu \lambda} 
     \partial_I x^\nu \partial_J x^\lambda, \tab
	G_{IJ} = s \partial_I x^\mu \partial_J x_\mu, \tab
	\partial_\mu \phi^I \partial_I x^\nu = {P_\mu}^{\nu}, \nonumber \\
	\delta G^{IJ} &= \big( s^{-1} \partial_\mu \phi^I \partial_\nu \phi^J 
   - \frac{1}{2} G^{IJ} P_{\mu \nu} \big) \delta g^{\mu \nu}, \tab
	u^\nu \nabla_\nu \nabla_\mu \phi = - \nabla_\mu u^\nu \nabla_\nu \phi,
  \label{formulas}
\end{align}
we find the stress takes precisely the form of the Hall viscosity
\begin{equation}
  \Delta T_{\mu \nu} = - \eta_H  u^\alpha \varepsilon_{\alpha \beta ( \mu} 
  {\sigma_{\nu)}}^\beta,
\end{equation}
where $\sigma_{\mu \nu} = {P_\mu}^{\alpha}{P_\nu}^{\beta} \big(
\nabla_\alpha u_\beta + \nabla_\beta u_\alpha - \frac{1}{2}
\nabla_\gamma u^\gamma P_{\alpha \beta}\big)$ is the shear tensor and
the Hall viscosity is $\eta_H = 2 s f$.  One may check that this gives a
nontrivial contribution to the tensor frame-invariant
\begin{align}
	C_T^{\mu \nu} &= P^{\mu \alpha} P^{\nu \beta} \Delta T_{\alpha \beta} - \frac{1}{2} P^{\mu \nu} P_{\alpha \beta} \Delta T^{\alpha \beta} = - \eta_H u_\alpha \varepsilon^{\alpha \beta ( \mu} {\sigma^{\nu )}}_\beta
\end{align}
and so cannot be removed by any field redefinition
\cite{Nicolis:2011ey,Haehl:2013kra}.

Moreover, $\eta_H$ represents a true response to geometric
perturbations.  Consider transverse, traceless perturbations about flat space with zero wavevector and
frequency $\omega$,
\begin{align}
	\delta g_{\mu \nu} &= 
	\begin{pmatrix}
		0 & 0 \\
		0 & \delta g_{ij} \
	\end{pmatrix},
	\tab \delta g_{ij} = h_{ij} e^{- i \omega_+ x^0},
	\tab h^i_{~ i} = 0,
\end{align}
where $\omega_+ = \omega + i \epsilon$ and $\epsilon \rightarrow 0^+$. These perturbations do not disturb the static solution $s =
\text{const}, u^\mu = ( 1,\, 0,\, 0 )$ to first order and the
stress-tensor two point function is easily evaluated.  We obtain the
traceless part of the long-wavelength Hall response
\begin{align}
	\eta^{\mu \nu \lambda \rho} (\omega ) &= \int \! d^3 x \, e^{i \omega_+ x^0}\Big( \frac{1}{i \omega_+} \left \langle \frac{\delta T_{\mu \nu} (x) }{\delta g_{\lambda \rho} (0)} \right \rangle + \frac{1}{2 \omega_+} \Theta ( x^0 ) \left \langle [ T^{\mu \nu} (x) , T^{\lambda \rho} (0 ) ]\right \rangle \Big) \nonumber \\
	&= \int \! d^3 x  \,  \frac{e^{i \omega_+ x^0}}{i \omega_+}  \frac{\delta \left \langle T^{\mu \nu} (x) \right \rangle}{\delta g_{\lambda \rho}(0)}= \eta_H u_\alpha \varepsilon^{\alpha \beta (\mu } P^{\nu ) (\lambda} P_\beta^{~ \rho )}
\end{align}
up to a contribution proportional to $g^{\mu\nu}g^{\lambda\rho}$ 
which is left undefined in this calculation.

\section{Number current and review of Friedman's formulation of fluid dynamics}

We now turn our attention to the case when the fluid has a conserved
charge.  Thus, we will expand our formalism to include other
independent currents.  One possible generalization involves
introducing a phase $\psi$ with a so-called chemical shift symmetry
$\psi \rightarrow \psi + c$ \cite{Dubovsky:2011sj}. The resulting
Noether current $n^\mu = n u^\mu$ is then interpreted as a conserved
particle density.  However, for our purposes this formalism is not
entirely suitable, as it breaks the symmetry between the conserved
entropy and particle number currents, leading to a less general form
of the Hall viscosity coefficient.
Instead, we will employ the formalism of Friedman
\cite{Friedman:1978wla}, used in his study of the stability of
relativistic stars, though in this discussion we principally follow
the treatment of Green, Schiffrin and Wald \cite{Green:2013ica}.

In Friedman's formalism, the $d$-dimensional space $\mathcal{F}$ is
replaced by a $(d+1)$-dimensional fiducial space $\mathcal{M}'$
diffeomorphic to the spacetime $\mathcal M$. The fluid configuration throughout time
is described by a diffeomorphism $\chi: \mathcal M \rightarrow
\mathcal M'$.  We will specialize to the case $d=2$.  $\mathcal
M'$ comes furnished with a fixed closed two-form $S_{AB}$ and a scalar
function $n'$ so that $N_{AB} = n'S_{AB}$ is also closed
\begin{equation}
	dS = d N  = 0.
\end{equation}
We denote $\mathcal M'$ indices with uppercase Latin letters near the
beginning of the alphabet $A$, $B$, $C$ etc.  The physical number and
entropy currents are then the pullbacks 
$S_{\mu \nu} = \chi^* (S_{AB})$ and
$N_{\mu\nu} = \chi^*( N_{AB} ) = n' S_{\mu \nu}$, respectively. 
The entropy and particle number currents $s^\mu = \frac{1}{2} \varepsilon^{\mu
  \nu \lambda} S_{\nu \lambda}$ and $n^\mu = n' s^\mu$ are identically
conserved and any given fluid configuration satisfying the two
conservation laws may be described by an appropriate choice of
$S_{AB}$ and $n'$.  We have thus restricted our description to include
only those fluid configurations that do not change the particle and
entropy content of a given fluid element
\begin{equation}
	\nabla_\mu s^\mu = 0,\tab
	\nabla_\mu n^\mu = 0 ,\tab
	u^\mu \nabla_\mu n' = 0 \tab
	\text{for all configurations}.
\end{equation}

Other relevant fluid variables are
\begin{align}
	s = (- s^\mu s_\mu)^{1/2}, \tab
	u^\mu = \frac{1}{s} s^\mu, \tab
	n = n' s,
\end{align}
and their transformations with respect to the metric are precisely as before
\begin{align}
	\delta s = - \frac12 s P^{\mu \nu} \delta g_{\mu\nu}, \tab
	\delta n' = 0, \tab
	\delta u^\mu = \frac12 u^\mu u^\nu u^\lambda \delta g_{\nu\lambda}.
\end{align}
If we set
\begin{align}
	S = - \int \! d^3x \, \sqrt{-g} \, \epsilon(s , n'),
\end{align}
we again obtain the perfect fluid stress tensor
\begin{align}
	T = \epsilon u^\mu u^\nu + P P^{\mu \nu},
\end{align}
where $P = s \partial_s \epsilon - \epsilon$.

To construct the WZW term above, consider the pushed forward metric
$g_{AB} = \chi_* ( g_{\mu \nu} )$ (we will always denote the
pushforward of a spacetime tensor by the same symbol but with Greek
indices replaced by upper case Latin indices).  The form
\begin{equation}
  \omega = - f(n') S_{AB} d g^{BC} g_{CD} \wedge d g^{DA} 
  \label{covariant_form}
\end{equation}
is then a closed two form on the space of Lorentzian metrics for any
function $f$ of the particle number-entropy ratio. Define $h$ to be
the orthogonal projector
\begin{equation}
  h_{AB} = g_{AB} + \frac{s_A s_B}{s^2}\,,
\end{equation}
then we also have
\begin{equation}
  \omega = \frac12 f(n') s^E \varepsilon_{EAB} d h^{BC} h_{CD}
            \wedge d h^{DA} .
\end{equation}

We may then recover Eq.~(\ref{omega-def}) by introducing coordinates
$\phi^I$, $I = 1,2$ on a spatial slice. Transport them along $u^A$,
and take $\phi^0$ to be the affine parameter to obtain a coordinate
patch on $\mathcal M'$. We call such a system ``comoving coordinates''
on $\mathcal M'$. $\omega$ then becomes
\begin{equation}
  \omega = \frac12 f(n') s \epsilon_{IJ} d H^{JK} H_{KL} \wedge d H^{LI},
\end{equation}
where $H_{IK}$ is $h_{IK}$ normalized to be unimodular, and
$\epsilon_{12} = 1$, as in the previous section.

Thus the two-form on $\mathbb H^2$ considered earlier may be
equivalently thought of as a (more obviously slicing independent)
two-form on the 6-dimensional space of Lorentzian metrics.  Working in
Friedman's formalism also buys us the added freedom to add an
arbitrary function of the number-entropy ratio since $n'$ is
identically conserved along worldlines. This is the reason we stick to this formalism. In the treatment common to the condensed matter literature involving chemical shifts \cite{Dubovsky:2011sj}, only the entropy current is identically conserved, whereas $n^\mu$ is merely conserved on shell. In the off shell formulation, $n'$ is not then merely a function of the fluid coordinates $\phi^I$ and promoting $f$ to $f (n' )$ would be nonsensical in (\ref{WZW coord}).

Writing down the WZW term in coordinates
\begin{align}
  S_{\rm WZW} = \frac{1}{2} \int \! d^3x \, du \, \sqrt{-g} f(n') S^\mu \Tr ( \epsilon \partial_u H^{-1} H \partial_\mu H^{-1}),
\end{align}
one may use this to explicitly check interpolation independence as in
Eq.~(\ref{total_derivative}).  Under an infinitesimal change of
interpolation, we find $\delta S$ is a total $u$ derivative plus a
term proportional to $\nabla_\mu ( f (n') s^\mu ) = 0$.

To complete the comparison with our original treatment, note that in the adapted choice of coordinates given above
\begin{equation}
   u^\mu \partial_\mu \phi^I = 0 ,\tab
   s^\mu = \frac12 \varepsilon^{\mu \nu \lambda} S_{IJ} 
      \partial_\nu \phi^I \partial_\lambda \phi^J, \tab
   n^\mu = \frac12 \varepsilon^{\mu \nu \lambda} n' S_{IJ} \partial_\nu \phi^I \partial_\lambda \phi^J .
\end{equation}
Through a $\phi^0$ independent redefinition of the spatial coordinates
$\phi^I$ we may further bring $N^\mu$ into the form
\begin{equation}
  s^\mu = \frac12 \varepsilon^{\mu \nu \lambda} 
    \epsilon_{IJ} \partial_\nu \phi^I \partial_\lambda \phi^J .
\end{equation}
If we wish to retain this formula, we may only perform area preserving
diffeomorphisms of the coordinates $\phi^I$.  We thus see our original
formalism for a single current arises in this setting via a suitable
choice of coordinates on the fluid manifold. Specifically, we choose
coordinates on $\mathcal M'$ that ``follow'' the paths of fluid
particles, which was our original interpretation of the scalars
$\phi^I$, plus an arbitrary normalization convention that restricts us
to area-preserving diffeomorphisms.

The variation of the action proceeds as before, but is much more
easily done in the form~(\ref{covariant_form}) since we simply have
$\delta g^{AB} = \partial_\mu \phi^A \partial_\nu \phi^B \delta g^{\mu
  \nu}$. We again obtain
\begin{align}
  \Delta T_{\mu \nu} &= \eta_H u^\alpha 
  \varepsilon_{\alpha \beta ( \mu} {\sigma_{\nu )}}^\beta,
\end{align}
where the Hall viscosity now has the more general form
$\eta_H=sf(n')$.

\section{Conclusions}

We have shown that the Hall viscosity can be incorporated into the
Lagrangian description of ideal fluids through a WZW term.  This term
gives rise to a nontrivial Hall viscosity, which has to have the form
$\eta_H = s f(n/s)$.

However, from standard hydrodynamic arguments based, in particular, on
the second law of thermodynamics, the Hall viscosity $\eta_H$ may be
an arbitrary function of $s$ and $n$ \cite{Haehl:2013kra}.  The
difficulty may be traced to the stringent restrictions that the
topological nature of a WZW term places on its form, only allowing us
prefactors that are functions of comoving thermodynamic
variables. Only one such quantity exists in general: the specific
entropy. Adopting Friedman's formalism, which restricts to fluid
configurations with comoving specific entropy, we can thus enhance our
description to include Hall viscosities not directly tied to the
entropy density, but can go no further. Whether it is possible to
obtain a completely arbitrary $\eta_H$ via an action principle remains
an open problem.  We note that the action of anomalous hydrodynamics
in (3+1) dimensions needs to be formulated on a Schwinger-Keldysh
contour~\cite{Haehl:2013hoa}.  It is possible that the most general
Hall viscosity requires the theory to be formulated on such a contour.

\acknowledgments

This work is supported, in part, by a Simons Investigator Grant from
the Simons Foundation and the ARO-MURI 63834-PH-MUR grant.  We would
like to thank Alberto Nicolis, Mukund Rangamani, Josh Schiffrin, and
Bob Wald for discussions.


\begin{thebibliography}{99}

\bibitem{Avron:1995fg} 
  J.~E.~Avron, R.~Seiler and P.~G.~Zograf,
  {\em Viscosity of quantum Hall fluids,}
  Phys.\ Rev.\ Lett.\  {\bf 75} (1995) 697.

\bibitem{Son:2013xra} 
  D.~T.~Son and C.~Wu,
  {\em Holographic Spontaneous Parity Breaking and Emergent Hall Viscosity and Angular Momentum,}
  arXiv:1311.4882 [hep-th].

\bibitem{Dubovsky:2005xd} 
  S.~Dubovsky, T.~Gregoire, A.~Nicolis and R.~Rattazzi,
  {\em Null energy condition and superluminal propagation,}
  JHEP {\bf 03} (2006) 025
  [hep-th/0512260].

\bibitem{Dubovsky:2011sj} 
  S.~Dubovsky, L.~Hui, A.~Nicolis and D.~T.~Son,
  {\em Effective field theory for hydrodynamics: thermodynamics, and 
the derivative expansion,}
  Phys.\ Rev.\ D {\bf 85} (2012) 085029.

\bibitem{Nicolis:2011ey} 
  A.~Nicolis and D.~T.~Son,
  {\em Hall viscosity from effective field theory,}
  arXiv:1103.2137 [hep-th].

\bibitem{Haehl:2013kra} 
  F.~M.~Haehl and M.~Rangamani,
  {\em Comments on Hall transport from effective actions,}
  JHEP {\bf 10} (2013) 074
  [arXiv:1305.6968].

\bibitem{Maciejko:2013dia} 
  J.~Maciejko, B.~Hsu, S.~A.~Kivelson, Y.~Park and S.~L.~Sondhi,
  {\em Field theory of the quantum Hall nematic transition,}
  Phys.\ Rev.\ B {\bf 88} (2013) 125137
  [arXiv:1303.3041].

\bibitem{Witten} 
  E.~Witten,
  {\em Global aspects of current algebra,}
  Nucl.\ Phys.\ B {\bf 223} (1983) 422.

\bibitem{Read} 
  B.~Bradlyn, M.~Goldstein and N.~Read,
  {\em Kubo formulas for viscosity: Hall viscosity, Ward identities, and the relation with conductivity,}
  Phys. Rev. B {\bf 86} (2012) 245309
  [arXiv:1207.7021].

\bibitem{Friedman:1978wla} 
  J.~L.~Friedman,
  {\em Generic instability of rotating relativistic stars,}
  Commun.\ Math.\ Phys.\  {\bf 62} (1978) 247.

\bibitem{Green:2013ica} 
  S.~R.~Green, J.~S.~Schiffrin and R.~M.~Wald,
  {\em Dynamic and thermodynamic stability of relativistic, perfect fluid 
  stars,}
  arXiv:1309.0177.

\bibitem{Haehl:2013hoa}
  F.~M.~Haehl, R.~Loganayagam and M.~Rangamani,
  {\em Effective actions for anomalous hydrodynamics,}
  arXiv:1312.0610.


\end{thebibliography}
\end{document}